\documentclass[11pt,a4paper]{article}
\usepackage[dvips]{graphicx}	
\usepackage{theorem}		
\usepackage{psfrag}		
\usepackage[sumlimits,nointlimits]{amsmath}
\usepackage{amssymb,mathrsfs}	
\usepackage{amsmath}
\usepackage{parskip}
\usepackage[round]{natbib}
\usepackage{a4wide}
\newtheorem{Proposition}{Proposition}[section]

\begin{document}
\title{Modelling Bonds \& Credit Default Swaps using a Structural Model with Contagion}
\author{Helen Haworth\thanks {The Nomura Centre for Mathematical Finance, OCIAM Mathematical Institute, Oxford University, 24-29 St Giles, Oxford, OX1 3LB, England. email: helen.haworth@linacre.oxon.org. This work is kindly supported by Nomura and the EPSRC. We are grateful to Sam Howison and Ben Hambly for helpful input and discussions.},\quad Christoph Reisinger$^{*}$ and William Shaw\thanks{King's College, The Strand, London, WC2R 2LS, England}}
\date{Revised version, June 2007}
\maketitle

\section*{Abstract}

This paper develops a two-dimensional structural framework for valuing credit default swaps and corporate bonds in the presence of default contagion. Modelling the values of related firms as correlated geometric Brownian motions with exponential default barriers, analytical formulae are obtained for both credit default swap spreads and corporate bond yields. The credit dependence structure is influenced by both a longer-term correlation structure as well as by the possibility of default contagion. In this way, the model is able to generate a diverse range of shapes for the term structure of credit spreads using realistic values for input parameters.

\section{Introduction}\label{intro}

Firms do not operate in isolation and company defaults are not independent. In reality a whole network of links exists between companies in related businesses, industries and markets and the impact of individual credit events can ripple through the market as a form of contagion. It is thus of fundamental importance when modelling credit, not only to understand the drivers of credit risk at an individual company, but also the dependence structure between related companies. Whether accounting for counterparty risk in the price of a single-name credit derivative, or considering credit risk in a portfolio context, an understanding of credit dependence is essential to accurate risk evaluation and pricing.

Credit weakness, ratings downgrades and ultimately corporate default can occur in three main ways. Firstly, a company may be adversely affected for reasons specific to that company alone (e.g.\ poor financial management). Secondly, credit weakness may occur due to a factor or factors impacting multiple companies -- whether in the form of a cyclical influence related to the economy, or a market-wide shock such as an earthquake or September 11$^{th}$. Finally, companies are related through ties, some of which are real (e.g.\ a trade-creditor agreement), others of which are purely a matter of perception -- for example the fear of accounting fraud. Credit dependence then occurs primarily through two mechanisms -- either as a direct consequence of a common driving factor, or due to inter-company ties. The latter can be thought of as a form of contagion.

The purpose of this paper is to examine the importance of considering a more complete dependence structure than is usually incorporated in credit models, one that better reflects reality, taking into account both a common driving influence and the possibility of idiosyncratic company links. To do so, whilst retaining some degree of analytical tractability, we consider a two-dimensional structural model with default as the first hitting time of an exponential barrier. Firm values are modelled as correlated geometric Brownian motions and the default event is contagious. In this way, we are able to capture the two facets of the dependence structure. The correlation in firm values reflects a longer-term common driving influence on corporate strength whilst default contagion represents a direct link between the fortunes of both companies. As discussed in Section \ref{bond}, this framework results in a model that is asymmetric with regard to default risk, a significant improvement on prior models.

Structural models, whilst far from straightforward mathematically, are far more grounded in economic fundamentals than many other models and thus form a good starting point for a realistic description of credit dynamics. \cite{Giesecke-2004}, \cite{Schonbucher} and \cite{Lando} provide a good introduction to structural models, their development since first introduced by \cite{Merton-1974}, and their traditional place within credit modelling. Until recently, the vast majority of work on the structural model has focused on the case of a single firm. Indeed, two popular commercial packages, Moody's KMV and CreditGrades$^{TM}$, are motivated by the single-firm structural model\footnote{Further details of these approaches can be found at www.moodyskmv.com and in \cite{Finger-2002}, respectively.}. Very little, however, has been published for multiple companies, with the market mainly focused on copulas or conditionally independent factor models\footnote{For a good overview of the use of copulas in finance, see \cite{Cherubini}; \cite{Schonbucher} provides a useful summary and references for factor models.} in the multivariate case. Two exceptions are the papers by \cite{Zhou-2001b} and \cite{Hull-2000b}. \cite{Zhou-2001b} calculates default correlations for two firms whose asset values are modelled as correlated Brownian motions. \cite{Hull-2000b} extend this to framework to higher dimensions and proceed numerically in a discrete-time setting, proposing a method to calibrate piecewise constant default barriers to a term structure of default hazard rates. In contrast, we extend the framework used in \cite{Zhou-2001b} to incorporate default contagion and derive analytical formulae for bond yields and CDS spreads.

Copula methods, which allow the dependence structure of a portfolio to be considered independently from individual default times, are easy to implement and have rapidly become the market standard for modelling portfolios of credits. However, as basically static models able only to model expected defaults over a given time period, they fail to allow for suitable credit spread dynamics and tend to exhibit time instabilities.\footnote{\cite{Mikosch-2005} provides a critical discussion of the widespread usage of copula methods; some countering arguments are given by \cite{Remillard}.} Furthermore, the copula approach has no notion of default cause and effect as exists in a contagion mechanism. 

As problems have arisen with the widespread market use of copulas, multidimensional structural models are seeing renewed interest. \cite{Hull-2005} price CDO tranches in a structural framework where assets are driven by a common factor. In this way, defaults are modelled in a dynamic setting, and firm value correlations can be time-dependent or stochastic, however the dependence structure stems purely from the correlated firm values and so is unable to account for any default causality or contagion. In another approach, \cite{Schoutens-2005}, \cite{Moosbrucker} and \cite{Baxter} assume that firm values are driven by Levy processes rather than geometric Brownian motions. By assuming that firm values are modelled as geometric Brownian motions time-changed by a common Gamma business time, firm values become Variance Gamma processes, allowing for a richer characterisation of spread dynamics. Dependence is introduced through having a common stochastic time change, which may then be further broken down into systematic and idiosyncratic components, with a number of different representations proposed by the various authors. The resultant dependence structure is more realistic and flexible but does not incorporate any form of default contagion. 

The paper is organised as follows. In Section \ref{model} we outline the framework for the model, its assumptions and underlying results. We consider both the most general formulation and cases in which the formulae simplify. Formulae and results for corporate bond yields are provided in Section \ref{bond}, whilst those for credit default swap spreads are in Section \ref{CDS}. We summarise and consider future extensions in Section \ref{conclusion}. Mathematical details are in the appendix.

\section{The Model}\label{model}

We consider two companies, firm values $V_i$, $i=1,2$. Each company issues equity and a single homogeneous class of debt, assumed to be a zero coupon bond, $C_i(t,T)$, par value $K_i$, maturity $T$. For simplicity we assume that both bonds have the same maturity date but the analysis is easily extendible to different maturity dates.

For each company, firm value is assumed to follow a geometric Brownian motion, with default as the first time that the value of the firm hits a lower default barrier $b_i(t)$. As in \cite{Merton-1974} and \cite{Black-1976}, we assume that a firm's value can be constructed from tradable securities and so in the risk-neutral pricing measure, for $i=1,2$,
\begin{displaymath}
dV_i(t) = (r_{\!f}-q_i)V_i dt + \sigma_i V_i dW_i(t)
\end{displaymath}
where the risk-free rate, $r_{\!f}$, dividend yields, $q_i$, and volatilities, $\sigma_i$, are constants, $W_i(t)$ are Brownian motions and cov$(W_1(t), W_2(t)) = \rho t$ for constant correlation $\rho$. 

We assume that each company has an exponential default barrier, reflecting the existence of debt covenants, and denote the default barrier for company $i$ by
\begin{displaymath}
b_i(t) = K_i e^{-\gamma_i(T-t)}.
\end{displaymath}
Whilst similar to the barrier formulation in \cite{Black-1976}\footnote{\cite{Black-1976} use a barrier of the form $\omega_i K_i e^{-r_{\!f}(T-t)}$ where $0 \leq \omega_i \leq 1$. The similarities and differences in this formulation compared to the one we use are discussed further in Section \ref{bond}.}, defining the barrier in this way provides us with additional flexibility, enabling us to change its slope. In particular, in the special case that the barrier growth rate is set equal to the drift in firm value, the model simplifies. As neither the barrier growth rate nor the drift in firm value is observable in practice, this is an attractive simplification. It is by no means necessary for our analysis, however, and we provide results for the most general case.

Setting   
\begin{displaymath}
X_i(t) = \ln \left(\frac{V_i(t)}{V_i(0)} e^{-\gamma_i t}\right)
\end{displaymath}
enables us to consider the simpler case of Brownian motion with drift and constant default barrier $B_i = \ln \left(\frac{b_i(0)}{V_i(0)}\right) \leq 0$. $X_i(0) = 0$ and

\begin{displaymath}
X_i(t) = \alpha_i t + \sigma_i W_i(t)
\end{displaymath}

where $\alpha_i = r_{\!f}-q_i-\gamma_i - \frac{1}{2} \sigma_i^2$.

Defining the running minimum
\begin{displaymath}
\underline{X}_i(t) = \min_{0 \leq s \leq t} X_i(s),
\end{displaymath}
and default time, $\tau_i$, as the first hitting time of the default barrier,
\begin{displaymath}
\tau_i = \inf \{t: X_i(t) = B_i \},
\end{displaymath}
survival probability is then
\begin{displaymath}
\mathbb{P}(\tau_i > s) = \mathbb{P}(\underline{X}_i(s) \geq B_i).
\end{displaymath}

The key result we use to value credit spreads is the joint survival probability density function and the resultant joint survival probability, $ P(t)$,

\begin{eqnarray}\label{eqn:survival}
P(t) &=& \mathbb{P}( \underline{X}_1(t) \geq B_1, \underline{X}_2(t) \geq B_2) \\
&=& \frac{2}{\beta t} e^{a_1B_1 + a_2B_2 + bt} \sum_{n=1}^{\infty} e^{-r_0^2/2t} \sin \left(\frac{n\pi\theta_0}{\beta}\right) \int_0^{\beta} \sin \left(\frac{n\pi\theta} {\beta}\right) g_n(\theta) \,\mathrm{d}\theta \nonumber
\end{eqnarray}
where
\begin{eqnarray*}
g_n(\theta) &=& \int_0^{\infty} r e^{-r^2/2t} e^{A(\theta)r} I_{(\frac{n\pi}{\beta})}\left(\frac{rr_0}{t}\right) \,\mathrm{d}r \\
a_1 &=& \frac{\alpha_1 \sigma_2 - \rho \alpha_2\sigma_1}{(1-\rho^2)\sigma_1^2\sigma_2}, \quad a_2 = \frac{\alpha_2 \sigma_1 - \rho \alpha_1\sigma_2}{(1-\rho^2)\sigma_1\sigma_2^2} \\
b &=& -\alpha_1 a_1 - \alpha_2 a_2 + \frac{1}{2} \sigma_1^2 a_1^2 + \rho \sigma_1 \sigma_2 a_1 a_2 + \frac{1}{2} \sigma_2^2 a_2^2 \\
\tan \beta &=& -\frac{\sqrt{1-\rho^2}}{\rho}, \quad \beta \in [0,\pi] \\
r_0 &=& \frac{1}{\sqrt{1-\rho^2}}\left( \frac{B_1^2}{\sigma_1^2} - \frac{2\rho B_1 B_2}{\sigma_1\sigma_2} + \frac{B_2^2}{\sigma_2^2} \right)^{1/2}  \\
\tan \theta_0 &=& \frac{\sigma_1 B_2 \sqrt{1-\rho^2}}{\sigma_2 B_1 - \rho \sigma_1 B_2}, \quad \theta_0 \in [0,\beta] \\
A(\theta) &=& a_1\sigma_1 \sin(\beta-\theta) + a_2\sigma_2\sin\theta,
\end{eqnarray*} 
and $I_{(\frac{n\pi}{\beta})} \left( \frac{rr_0}{t} \right)$ is a modified Bessel's function. $P(t)$ represents the probability that neither company hits its default barrier by time $t$. Correlation between firm values is reflected by $\beta$, with $\beta=\pi$ corresponding to perfect correlation, $\beta=\pi/2$ independence and $\beta=0$ perfect negative correlation. When the drifts in firm value and the default barrier are equal, $\alpha_i=0$, leading to $a_i=0=b$. The form of $P(t)$ derives from the separation of variables in the solution of the Fokker-Planck equation governing the evolution of the survival probability density function. Full details of the calculation methodology can be found in the paper by \cite{Rebholz} in which double lookbacks are valued using the joint distributions for the maxima and minima of two correlated Brownian motions. \cite{Zhou-2001b} uses the same result to calculate default correlation for two companies with correlated assets. In particular, Zhou\footnote{N.B. The definitions of $a_1$ and $a_2$ in Zhou are the negative of the definitions in this paper. Our formula for $g_n(\theta)$ can be reconciled with Zhou's using double angle formulae and the fact that $\cos \beta = -\rho$, $\sin \beta = \sqrt{1-\rho^2}$. (His angle $\alpha$ is our angle $\beta$.)} primarily considers the special case that the barrier growth rate and drift in firm value are equal, $\gamma_i = r_{\!f}-q_i-\frac{1}{2} \sigma_i^2$, and shows that

\begin{equation}\label{eqn:nodrift}
P(t) = \frac{2r_0}{\sqrt{2 \pi t}}e^{-r_0^2/4t} \sum_{n=1,3,...}\frac{1}{n} \sin \left(\frac{n\pi\theta_0}{\beta}\right) \left[ I_{\frac{1}{2}(\frac{n\pi}{\beta}+1)}\left(\frac{r_0^2}{4t}\right) +  I_{\frac{1}{2}(\frac{n\pi}{\beta}-1)}\left(\frac{r_0^2}{4t}\right) \right].
\end{equation}

This simplification, implying constant leverage, makes the joint survival probability considerably faster to evaluate. \cite{Zhou-2001b} found that the assumption had minimal impact on default correlations for shorter maturities. Supporting this finding we see very little difference in survival probabilities as the barrier growth rate, $\gamma_i$, is changed for maturities up to five years. This is reflected in the fact that implied bond yields are not very sensitive to changes in the barrier growth rate, as illustrated in Figure \ref{fig:bondw7gamma}.\footnote{An extensive analysis of survival probability sensitivity to input parameters is contained in \cite{Haworth_DPhil}; here we consider solely the impact of key parameter assumptions with regard to credit spreads, with results in Section \ref{bond}.}

The joint survival probability calculation can also be simplified by selecting values of $\beta$ for which the modified Bessel's function simplifies -- for example $\beta=\frac{\pi}{k}$ for integer $k$. Unfortunately, since $\rho = -\cos \beta$, the majority of these cases correspond to negative values of correlation and so are of only limited interest in our framework.

\section{Bond Yield Calculation}\label{bond}

The value of a corporate bond to a bondholder arises from two components -- its value on maturity (should it mature) and its value in the event of default. We consider the $t=0$ value of a zero coupon bond issued by company one, par value $K_1$, maturing at time $T$, denoted $C_1(0,T)$. The yield is then

\begin{equation}\label{eqn:yieldformula}
y_1(0,T) = -\frac{1}{T} \ln \left( \frac{C_1(0,T)}{K_1} \right).
\end{equation}

We incorporate default contagion by assuming that company one defaults on its outstanding debt the first time that the value of either company reaches its default barrier. In this way, there are two components to the default mechanism. If the value of firm one declines sufficiently, the company is forced into bankruptcy -- this is due to the direct performance of the company itself and is exactly the framework used in normal first-passage structural models. The second way in which company one can default is if company two goes bankrupt, modelled as the time when the value of company two reaches its default barrier -- default contagion. This would be the situation if company two was essential to the continuing operation of company one. For example if company two was the only purchaser of company one's products, as in the case of a small or regional auto-parts supplier to General Motors. 

Our specification of the contagion mechanism implicitly assumes that the full extent of its existence is not known to the market beforehand,\footnote{We are grateful to one of our referees for highlighting this.} as is so often the case. Many links between companies are very opaque and certainly not fully disclosed. For example, the full nature of a bank's loan portfolio is rarely apparent, and there are many instances when banks have ended up over-exposed to one company or industry. The situations at Parmalat and LTCM in recent years are cases in point. Another example is the networks of business links that exist, for example in Italy, when extensive corporate cross-holdings are common. Few of these are publicly known. Too frequently, the existence or full extent of a company's relationships and the consequent investment risks only come to light when problems arise. The goal of our model is to take the first step in considering the spread implication of these types of links through the introduction of a contagion process that is simple enough to allow for analytical solutions.        
It is worth noting that company two need not default automatically if company one does. It can continue to operate regardless of the financial viability of company one with dependence solely through the asset correlation, $\rho$. As a result, the model is asymmetric with respect to default risk, in stark contrast with the majority of previous models incorporating a credit dependence structure.\footnote{The natural extension of this framework to larger portfolios of companies would be the type of situation considered in \cite{Jarrow-2001} in which `primary' companies impact `secondary' companies but not vice versa. A `primary' company is likely to be larger with a greater market impact than a `secondary' company. For example Microsoft or General Motors compared to a small, local IT or auto component manufacturer.}

This framework is only really realistic for $\rho \geq 0$. It is highly unlikely that the bankruptcy of one company would lead to the immediate default of another, negatively correlated company. Whilst it is possible to contrive a theoretical example (e.g. a highly diversified company like General Electric might be key to one of its suppliers, but negatively correlated with it overall), economically it is rather improbable in practice and so we restrict ourselves to consideration of $0 \leq \rho \leq 1$.

Similarly to \cite{Black-1976}, we assume that 
\begin{displaymath}
\mbox{Payment at maturity} = \min (\omega_1 V_1(T), K_1) \quad \mbox{provided} \quad \tau_1>T, \tau_2>T
\end{displaymath}
where $K_1$ is the par value of the bond, $\tau_i$ denotes the default time of company $i$, and $\omega_1$ is a constant write down factor. This factor is the same as used later in the specification of the payment on default. It represents the fact that in the event of default or a restructuring, a portion of the defaulting company's value is lost to bondholders. This is commonly seen in practice, caused, for example, by restructuring costs or violation of the priority rule allocating claims in the event of default.

\cite{Black-1976} incorporate the write-down factor in the default barrier, but then the bondholders are always paid in full once the value of the firm at maturity is at least the face value of the bonds. This seems a little unrealistic since a company would be unable to pay out its entire value to bondholders. There would be liquidation costs and a company would not be able to raise its entire value in a refinancing. It therefore makes sense that the firm value at maturity must be equal to some $K_1/\eta_1$, where $\omega_1 \leq \eta_1 \leq 1$, for bondholders to be fully repaid. For simplicity, since fewer parameters are preferable, we assume that the firm repays bondholders in full for $V_1(T) \geq K_1/\omega_1$. This would seem to be an improvement on the case where full repayment occurs for firm value at maturity of par or more.

Changing to $X_i(t)$ coordinates, the discounted maturity payment, DMP, can be written
\begin{eqnarray*}
\textrm{DMP} &=& e^{-r_{\!f}T}  \int_{B_2}^{\infty} \int_d^{\infty} K_1 p(x_1, x_2, T) \,\mathrm{d}x_1 \,\mathrm{d}x_2\\
&& {}+  e^{-r_{\!f}T}  \int_{B_2}^{\infty} \int_{B_1}^d \omega_1 V_1(0) e^{x_1 + \gamma_1 T} p(x_1, x_2, T) \,\mathrm{d}x_1 \,\mathrm{d}x_2 
\end{eqnarray*}
where 
\begin{equation}
p(x_1, x_2, T) = \frac{\partial^2}{\partial x_1 \partial x_2}\mathbb{P}(X_1(T) \leq x_1, X_2(T) \leq x_2, \underline{X}_1(T) \geq B_1, \underline{X}_2(T) \geq B_2) 
\end{equation}
is the joint survival probability density function at maturity and
\begin{displaymath}
d = \ln{ \frac{K_1}{\omega_1 V_1(0)}} - \gamma_1 T = B_1 - \ln \omega_1 \geq B_1.
\end{displaymath}

As outlined in Appendix \ref{app:1}, integrating and making a change of variables,
\begin{eqnarray}\label{maturitypayment}
\textrm{DMP} &=& H_1(T) \sum_{n=1}^{\infty} \sin \left(\frac{n\pi\theta_0}{\beta}\right) \int_0^{\beta} \sin \left(\frac{n\pi\theta} {\beta}\right) g_n^+(\theta) \,\mathrm{d}\theta\\
& & {}+ H_2(T) \sum_{n=1}^{\infty} \sin \left(\frac{n\pi\theta_0}{\beta}\right) \int_0^{\beta} \sin \left(\frac{n\pi\theta} {\beta}\right) g_n^{*}(\theta) \,\mathrm{d}\theta \nonumber
\end{eqnarray}
where
\begin{eqnarray*}
H_1(T) &=& \frac{2K_1e^{-r_{\!f}T}}{\beta T} e^{a_1B_1 + a_2B_2 + bT} e^{-r_0^2/2T}\\
H_2(T) &=& \frac{2\omega_1V_1(0)e^{(\gamma_1-r_{\!f})T}}{\beta T} e^{(a_1+1)B_1 + a_2B_2 + bT} e^{-r_0^2/2T}\\
g_n^+(\theta) &=& \int_{d^*(\theta)}^{\infty} r e^{-r^2/2T} e^{A(\theta)r} I_{(\frac{n\pi}{\beta})}\left(\frac{rr_0}{T}\right) \,\mathrm{d}r \\
g_n^{*}(\theta) &=& \int_{0}^{d^*(\theta)} r e^{-r^2/2T} e^{[A(\theta) + \sigma_1\sin(\beta-\theta)]r} I_{(\frac{n\pi}{\beta})}\left(\frac{rr_0}{T}\right) \,\mathrm{d}r \\
d^*(\theta) &=& \frac{d-B_1}{\sigma_1\left[\sqrt{1-\rho^2}\cos \theta + \rho \sin \theta \right]} = \frac{\ln \omega_1}{\sigma_1\sin(\theta-\beta)} \geq 0.
\end{eqnarray*}

We assume that default occurs the first time that either company hits its default barrier and that in the event of default, the bondholder receives a percentage of discounted par value, $\omega_1 K_1e^{-r_{\!f}(T-\tau)}$. This is the same payoff as used by \cite{Black-1976} and is highly attractive since the discounted default payment then becomes
\begin{equation}\label{defaultpayment}
\omega_1 K_1 e^{-r_{\!f} T} (1-P(T)),
\end{equation}
where the joint survival probability $P(T) = \mathbb{P}(\underline{X}_1(T) \geq B_1,\underline{X}_2(T)\geq B_2)$ is defined in (\ref{eqn:survival}).

\begin{figure}[htbp]
\setlength{\unitlength}{1cm}
\begin{minipage}[t]{8.0cm}
\caption{Implied yield curve, $\omega_1 = 0.7$}\label{fig:bondw7yc}
\includegraphics[width=8cm]{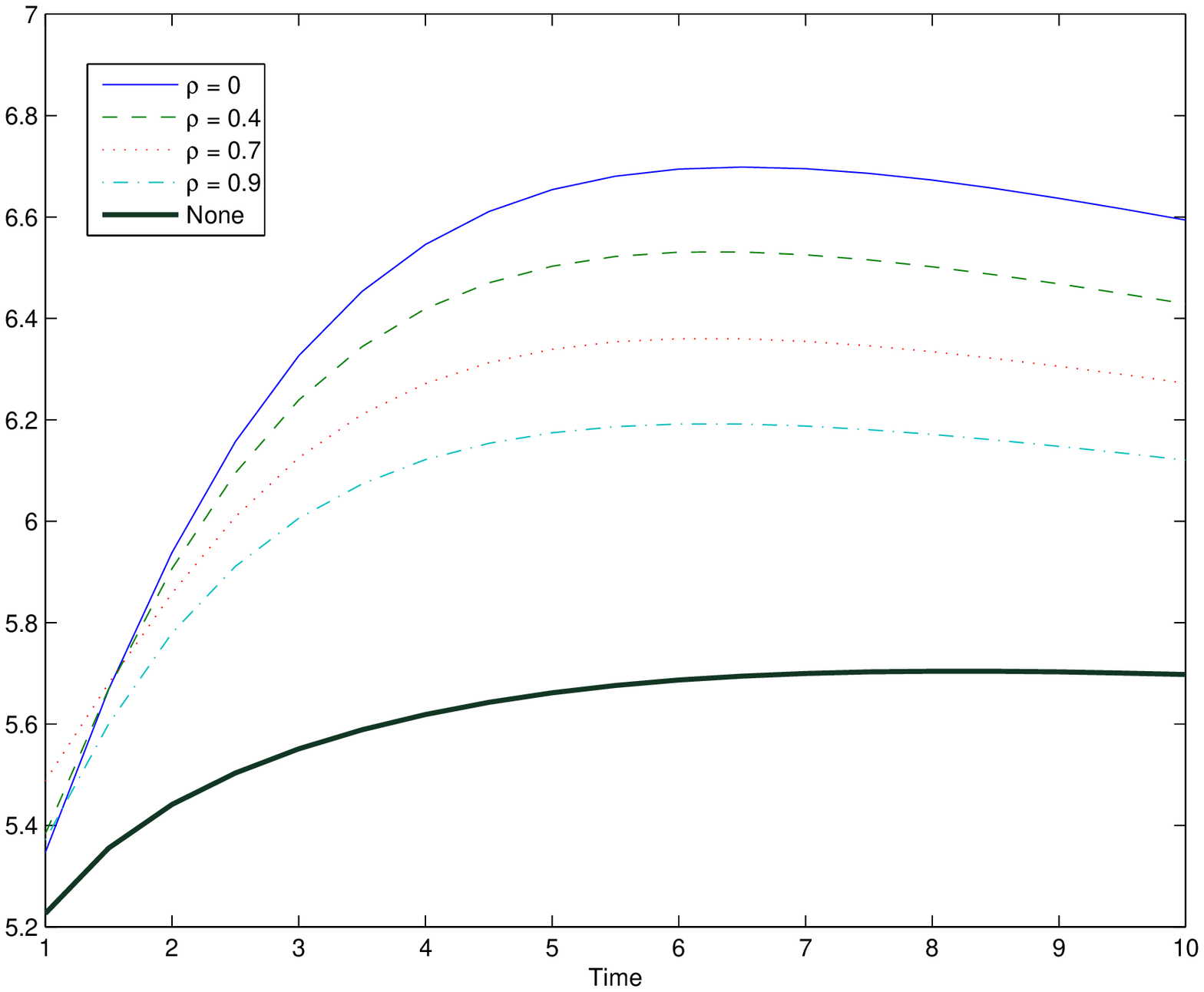} 
\end{minipage}
\hfill
\begin{minipage}[t]{8.0cm}
\caption{Implied yield curve, $\omega_1 = 0.5$}\label{fig:bondw5yc}
\includegraphics[width=8cm]{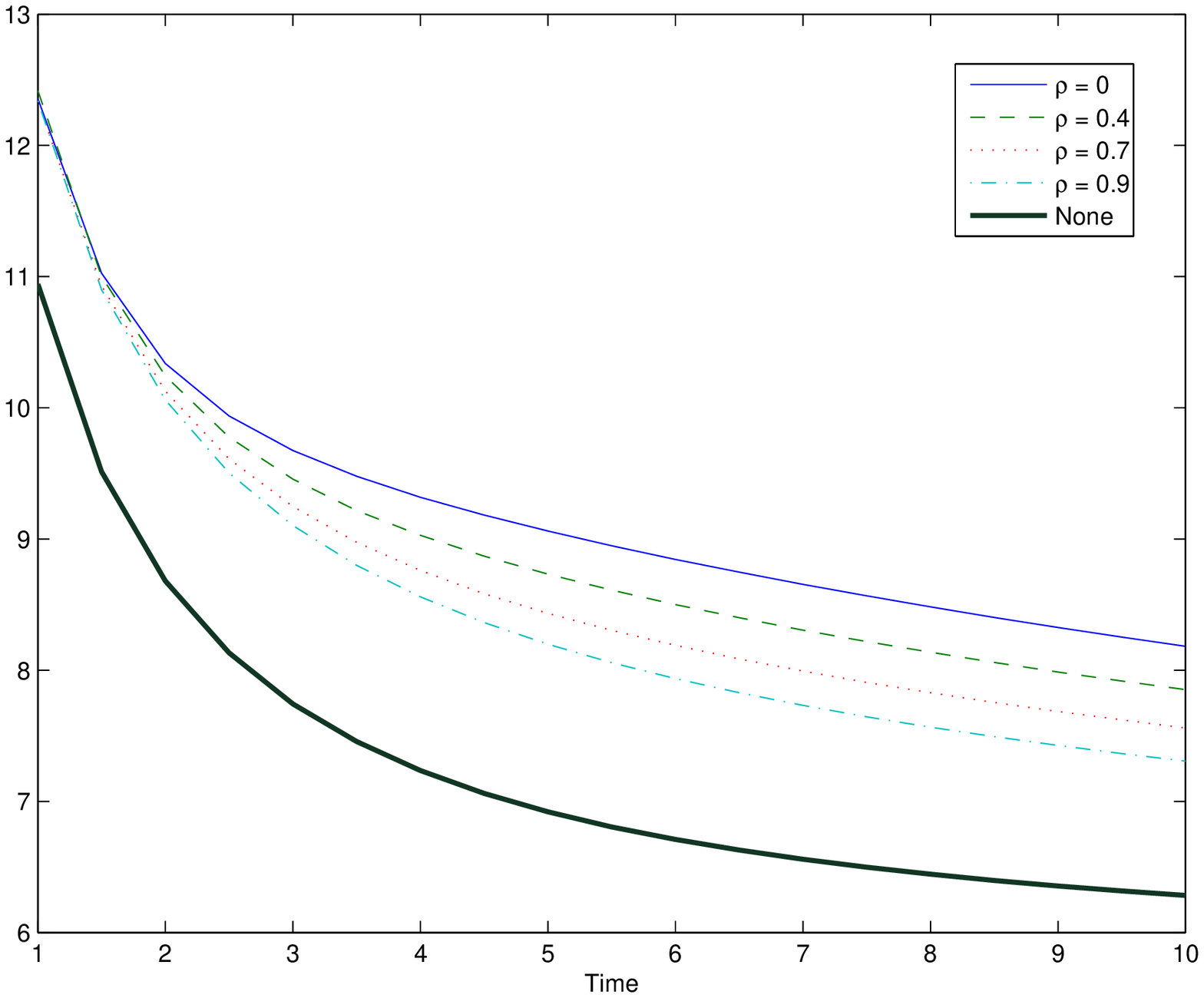}
\end{minipage}
\begin{center}
\small{$ \sigma_1=\sigma_2=0.2, K_1=100, r_{\!f}=0.05,$\\$ q_1=q_2=0, \gamma_1=\gamma_2=0.03$, initial credit quality = 2}
\end{center}
\end{figure}

\begin{figure}[htbp]
\setlength{\unitlength}{1cm}
\begin{minipage}[t]{8.0cm}
\caption{Implied bond yield, varying $\omega_1$}\label{fig:bondomega}
\includegraphics[width=8cm]{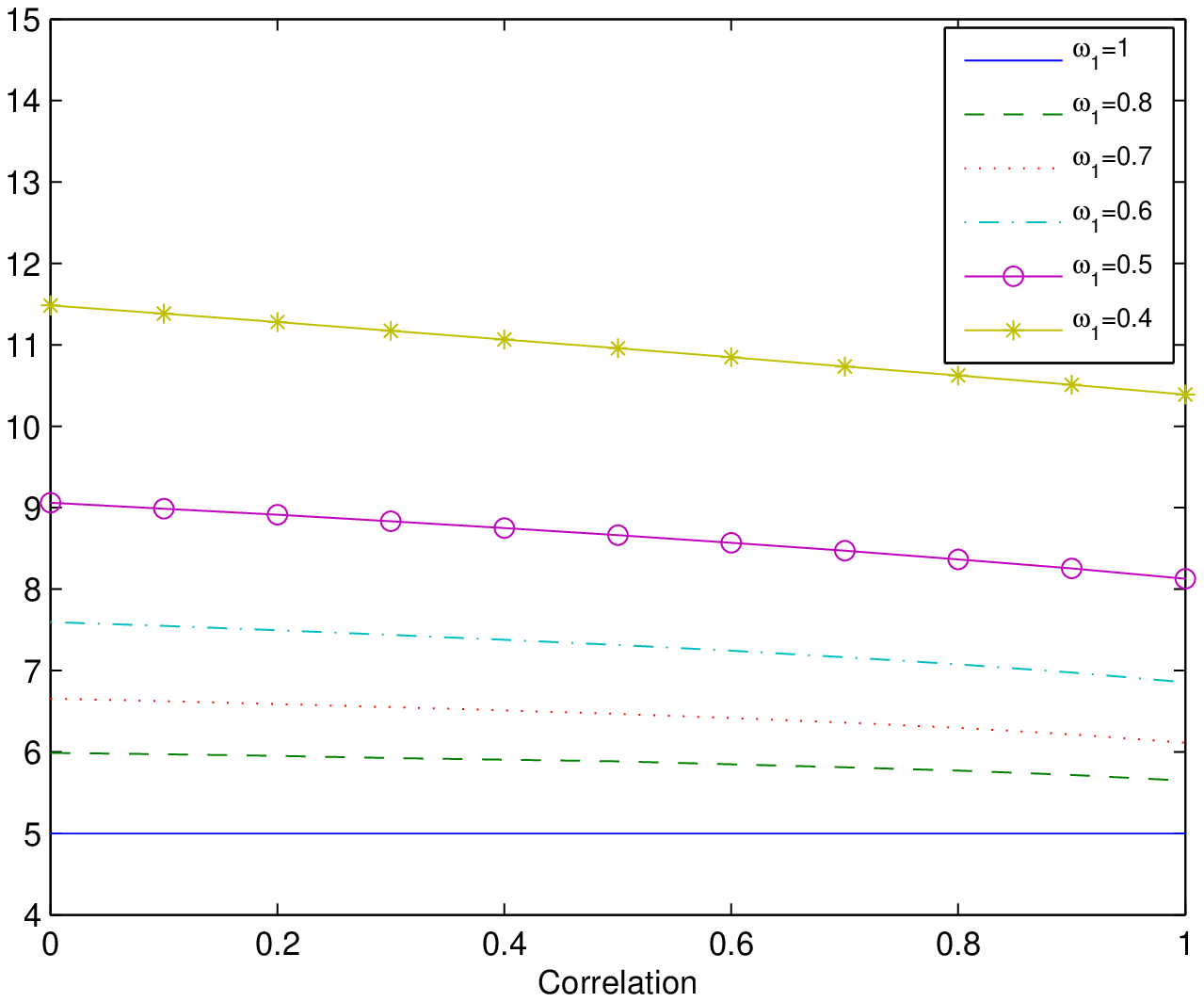}
\begin{center} 
\small{$ \sigma_i=0.2, K_1=100, r_{\!f}=0.05, q_1=q_2=0,$\\$  \gamma_i=0.03$, initial credit quality = 2, T=5}
\end{center}
\end{minipage}
\hfill
\begin{minipage}[t]{8.0cm}
\caption{Implied bond yield, varying $\gamma_i$}\label{fig:bondw7gamma}
\includegraphics[width=8cm]{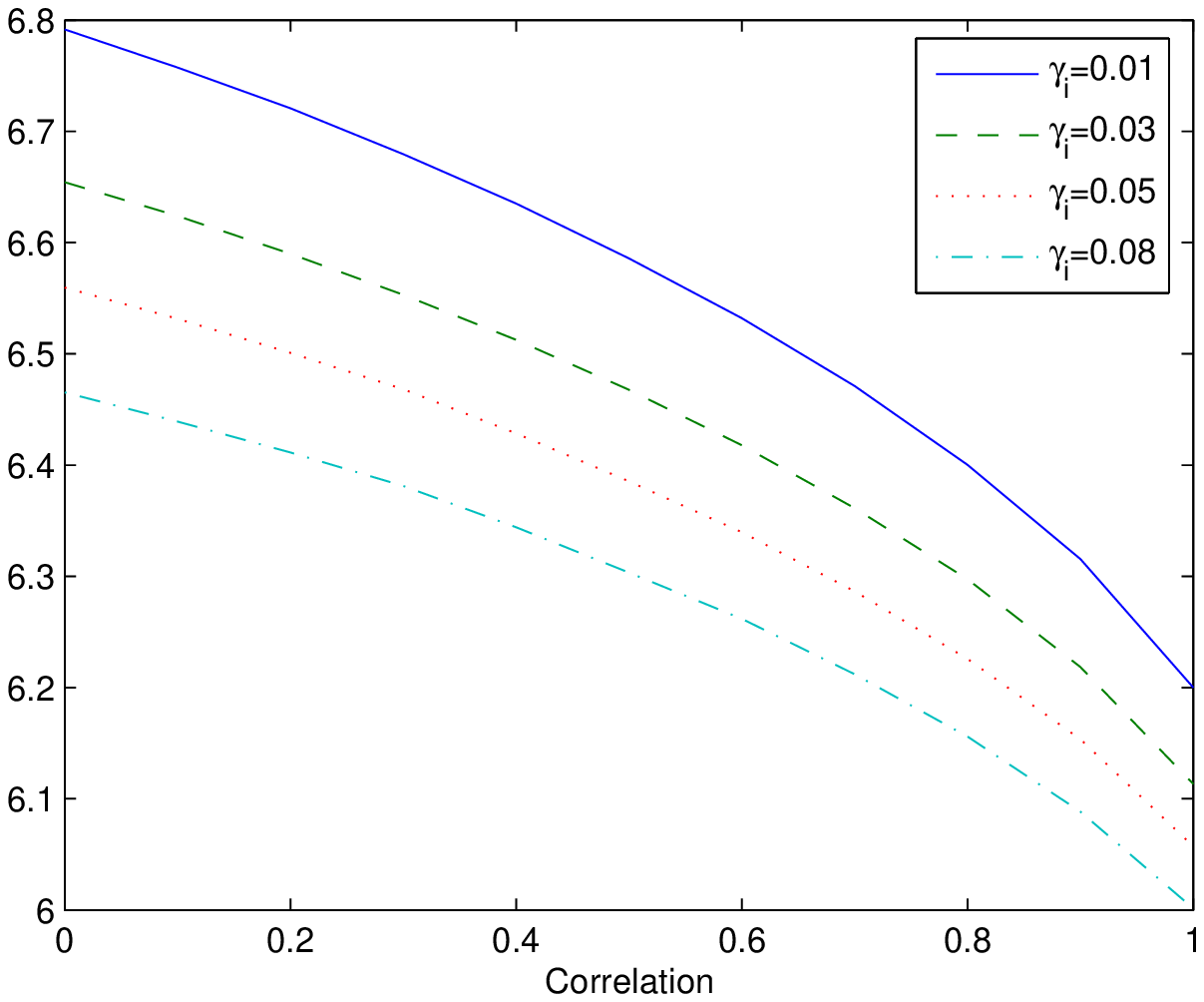}
\begin{center}
\small{$ \sigma_i=0.2, K_1=100, r_{\!f}=0.05,\omega_1 = 0.7,$\\$ q_i=0$, initial credit quality = 2, T=5}
\end{center}
\end{minipage}
\end{figure}

\begin{figure}[htbp]
\setlength{\unitlength}{1cm}
\begin{minipage}[t]{8.0cm}
\caption{Implied bond yield, varying $\sigma_i$}\label{fig:bondw7sigma}
\includegraphics[width=8cm]{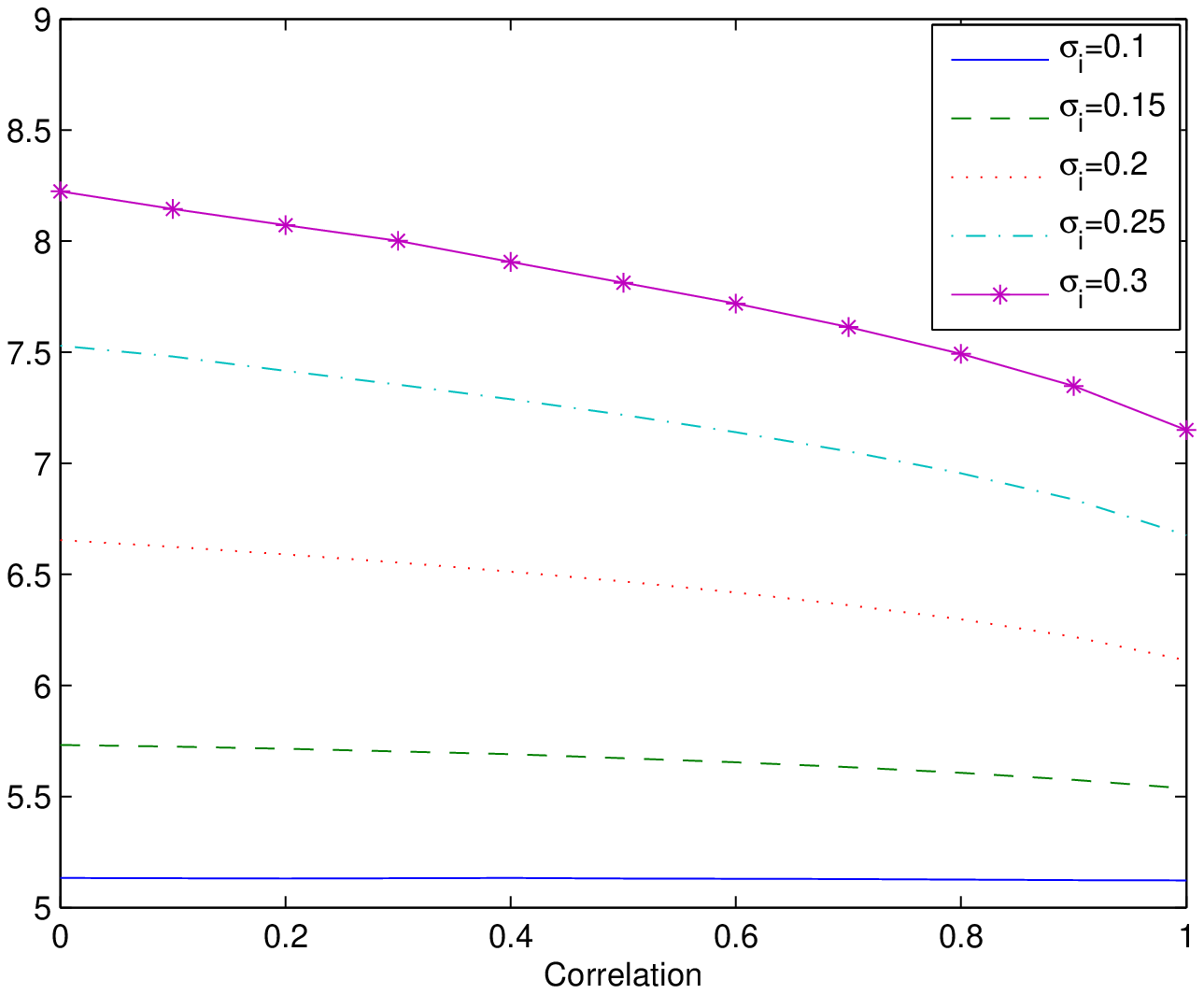} 
\end{minipage}
\hfill
\begin{minipage}[t]{8.0cm}
\caption{Implied bond yield, varying $\sigma_2$}\label{fig:bondw7sigma2}
\includegraphics[width=8cm]{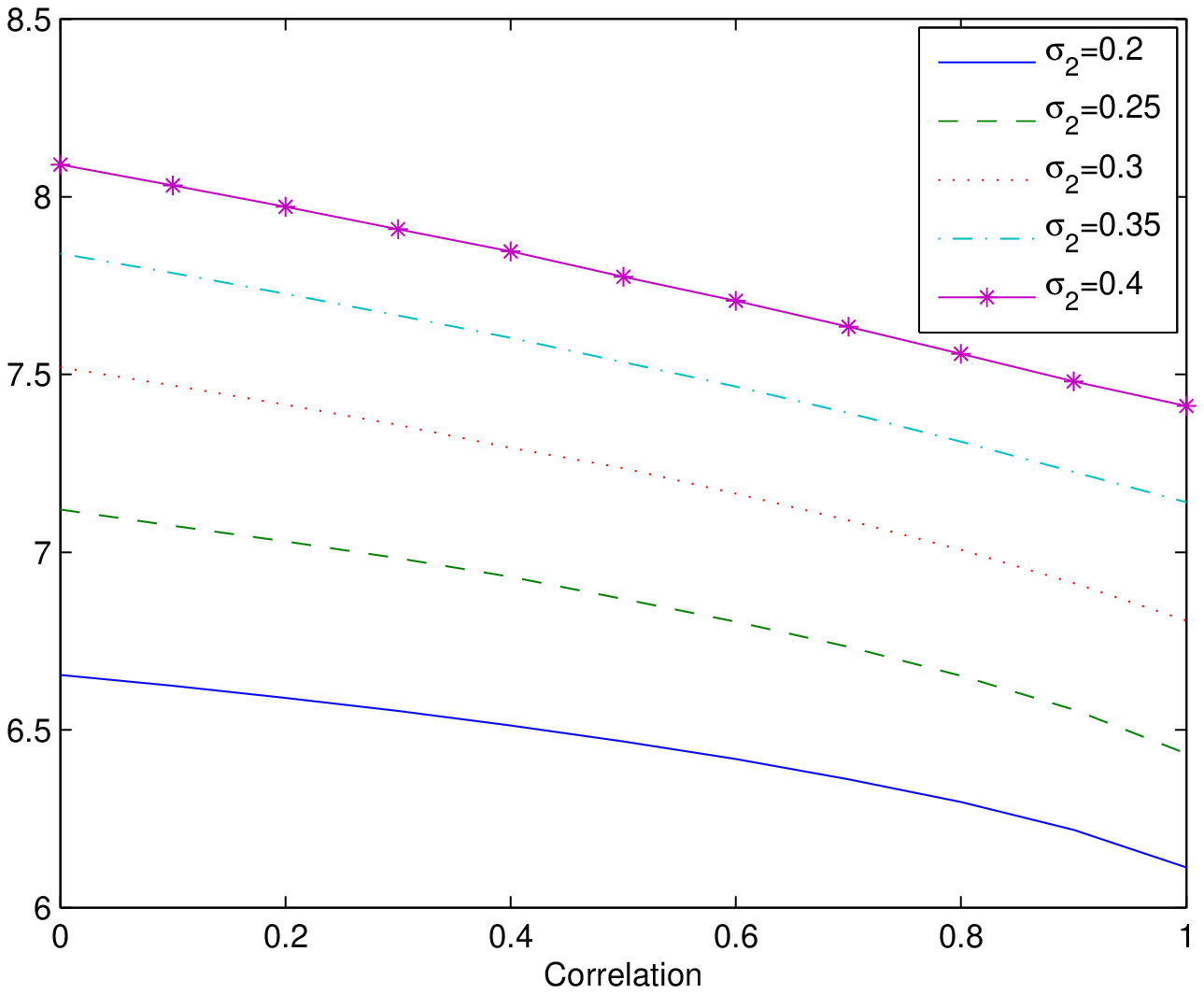}
\end{minipage}
\begin{center}
\small{$ \gamma_1=\gamma_2=0.3, K_1=100, r_{\!f}=0.05,\omega_1 = 0.7,$\\$ q_1=q_2=0$, initial credit quality = 2, T=5 years}
\end{center}
\end{figure}

By construction, the default payment is worth less than discounted par, and so for consistency we just need to ensure that it is worth less than the value of the firm at default. Since company one must be worth at least as much as its default barrier, a sufficient condition is  
\begin{displaymath}
\omega_1 \leq e^{(r_{\!f}-\gamma_1)(T-\tau_1)}.
\end{displaymath}
Adding together (\ref{maturitypayment}) and (\ref{defaultpayment}), and using (\ref{eqn:survival}) for $P(T)$, yields can then be calculated using (\ref{eqn:yieldformula}). In order to evaluate (\ref{maturitypayment}) and (\ref{defaultpayment}) we use the integral form of the modified Bessel's function,
\begin{equation*}
I_{(\frac{n\pi}{\beta})}\!\left(\frac{rr_0}{t}\right)\! = \frac{1}{\pi}\int_0^{\pi}\!\! e^{ \frac{rr_0}{t} \cos \phi} \cos \!\left(\frac{n\pi\phi}{\beta}\right)\!\,\mathrm{d}\phi - \frac{1}{\pi} \sin\!\left(\frac{n\pi^2}{\beta}\right)\!\int_0^{\infty}\!\!e^{-\frac{rr_0}{t}\cosh s - \frac{n\pi s}{\beta}} \,\mathrm{d}s.
\end{equation*}
The infinite sums in (\ref{maturitypayment}) and (\ref{defaultpayment}) converge rapidly to zero, and so this substitution enables us to approximate solutions by a finite sum of three-dimensional integrals which we evaluate by numerical quadrature using a sparse grid (for further information regarding sparse grid methods, see \cite{Gerstner-1998}). Figures \ref{fig:bondw7yc} - \ref{fig:bondw7sigma2} illustrate results and parameter sensitivities. As a measure of company strength we consider initial firm value, $V_i(0)$, divided by the initial level of the barrier, $b_i(0)$. We denote this parameter by initial credit quality and spread sensitivity to it is as would be expected -- as initial credit quality declines, spreads widen significantly. An indicative scaled distance to default, $\frac{1}{\sigma_i}\log(\frac{V_i(0)}{b_i(0)})$, for our parameter values of $\sigma_i=0.2$ and initial credit quality of 2 is then 3.5. 

In all cases, we see that yields decline as correlation increases. Since default is less likely with increasing correlation,\footnote{This is the case whether or not there is default contagion and can be seen from the fact that the probability of at least one of the companies defaulting in a given time period decreases as they become more correlated. If $N_i$ denotes default by company $i$, for $i=1,2$, then $\mathbb{P}(N_1 \cup N_2) = \mathbb{P}(N_1)+\mathbb{P}(N_2)-\mathbb{P}(N_1 \cap N_2)$ and the final term increases with $\rho$.} the bond is less risky, bond-holders are not rewarded with such high returns, increasing the price and reducing the yield. 

Figures \ref{fig:bondw7yc} and \ref{fig:bondw5yc} illustrate the yield curve for two values of the write-down factor, $\omega_1$. Comparing them, we see that as $\omega_1$ decreases, the yield curve inverts. The extent to which payments are written down clearly has a large impact on yields, particularly at the short-end of the yield curve. The thick black line, labelled `none' in the figures is the bond yield for a firm with the same parameter values, but operating in isolation. In other words, a single firm's yield when modelled in a first-passage framework, but with no exposure to another company through default contagion. As one would expect, yields are lower as such a firm's bonds are less risky. It is immediately apparent that allowing for default contagion has a significant impact on yields, especially for longer-dated bonds.

 Figure \ref{fig:bondomega} shows the impact of varying $\omega_1$ on a five-year bond. Of note, the case when $\omega_1 = 1$ corresponds to no write-down on default and in effect the bond becomes risk-free, yielding, as we would expect, the risk-free rate of $5\%$ regardless of correlation. $\omega_1$ acts in two ways -- it lowers the payment in the event of default and it increases the value the company must be worth at maturity for bondholders to be repaid in full.

Figure \ref{fig:bondw7gamma} considers the sensitivity of yields to the shape of the default barrier for a 5-year bond with initial credit quality of two and $\omega_1 = 0.7$. Changing the slope of the default barrier has minimal impact on yields -- as the slope increases, default is less likely and yields decrease, but the impact is fairly small, particularly when considering the dependence on other parameters. $\gamma_i$ has even less impact for shorter maturities, but becomes progressively more important as time to maturity increases.

Finally, we consider the impact of varying the volatility of firm-value. Figure \ref{fig:bondw7sigma} shows how yields behave when the volatility of both firms is changed simultaneously. As expected, higher volatility leads to a higher likelihood of default and higher yields. In Figure \ref{fig:bondw7sigma2} we assume that the volatility of firm one remains fixed at $0.2\textrm{ yr}^{-1/2}$ and we increase the volatility in firm two. Since we are considering the yield on firm one's bonds, the increasing riskiness of firm two impacts yields through the correlation between the two companies and the possibility of default contagion. As expected, the more volatile firm two is, the riskier firm one and the higher yielding its bonds.

\section{CDS Spread Calculations}\label{CDS}

Using a similar approach to that in Section \ref{bond}, we evaluate first and second-to-default credit default swap (CDS) spreads for a two-company basket.

\subsection{First-to-default CDS Basket}\label{basket}

We consider a basket of two related companies. The buyer of a first-to-default CDS on this underlying basket pays a premium, the CDS spread, for the life of the CDS -- until maturity or the first default, whichever happens first. In the event of default by one of the underlying reference companies, the buyer receives a default payment and the contract terminates. Denoting the default time of company $i$ by $\tau_i$, we write $\tau_{\textrm{first}}$ for the time of the first default, 
\begin{displaymath}
\tau_{\textrm{first}} = \min \{\tau_1, \tau_2\}
\end{displaymath}
where, using the same notation as before, 
\begin{displaymath}
\tau_i = \inf \{t: X_i(t) = B_i \}.
\end{displaymath}

If bond recovery on default is $R$, and the protection buyer makes continuous spread payments, $c$, on a par value $K$, then the discounted spread payment (DSP) and discounted default payment (DDP) on the first-to-default basket are
\begin{eqnarray}
\textrm{DSP } &=& cK\int_0^T e^{-r_{\!f}s}\mathbb{P}(\tau_{\textrm{first}}>s)\,\mathrm{d}s\nonumber \\
\textrm{DDP} &=& (1-R)K \int_0^T e^{-r_{\!f}s}\mathbb{P}(s \leq \tau_{\textrm{first}} \leq s+ ds )\,\mathrm{d}s  \label{cds1default}\\
&=& (1-R)K \int_0^T -e^{-r_{\!f}s}\frac{\partial}{\partial s}\mathbb{P}(\tau_{\textrm{first}} > s )\,\mathrm{d}s\nonumber \\
&=& (1-R)K \bigg \{1-e^{-r_{\!f}T}\mathbb{P}(\tau_{\textrm{first}}>T) - r_{\!f}\int_0^T e^{-r_{\!f}s}\mathbb{P}(\tau_{\textrm{first}} > s )\,\mathrm{d}s\bigg \}.\nonumber
\end{eqnarray}

With $P(s)$ defined as in equation (\ref{eqn:survival}), the market spread, $c_{\textrm{first}}$, is therefore
\begin{eqnarray}\label{eqn:cdsfirst}
c_{\textrm{first}} &=& \frac{(1-R) \bigg\{1-e^{-r_{\!f}T}\mathbb{P}(\tau_{\textrm{first}} > T )- \int_0^Tr_{\!f}e^{-r_{\!f}s}\mathbb{P}(\tau_{\textrm{first}} > s ) \,\mathrm{d}s \bigg \}}{\int_0^T e^{-r_{\!f}s}\mathbb{P}(\tau_{\textrm{first}} > s )\,\mathrm{d}s}\\ \nonumber
&=& \frac{(1-R) \bigg\{1-e^{-r_{\!f}T}P(T)- \int_0^Tr_{\!f}e^{-r_{\!f}s}P(s) \,\mathrm{d}s \bigg \}}{\int_0^T e^{-r_{\!f}s}P(s)\,\mathrm{d}s}
\end{eqnarray}
since 
\begin{displaymath}
\mathbb{P}(\tau_{\textrm{first}} > s )= \mathbb{P}(\tau_1>s, \tau_2>s) = \mathbb{P}(\underline{X}_1(s) \geq B_1,\underline{X}_2(s) \geq B_2)= P(s).
\end{displaymath}

\subsection{Second-to-default CDS Basket}\label{basket2}

A second-to-default CDS spread is evaluated in the same way. The purchaser of the swap receives a payment in the event that both companies default during the life of the swap, at which point the contract terminates. Denoting $\tau_{\textrm{second}}$ as the time of the second default, exactly as for (\ref{eqn:cdsfirst}), the market spread, $c_{\textrm{second}}$ is

\begin{equation}\label{eqn:cdssecond}
c_{\textrm{second}} = \frac{(1-R) \bigg\{1-e^{-r_{\!f}T}\mathbb{P}(\tau_{\textrm{second}} > T )- \int_0^Tr_{\!f}e^{-r_{\!f}s}\mathbb{P}(\tau_{\textrm{second}} > s ) \,\mathrm{d}s \bigg \}}{\int_0^T e^{-r_{\!f}s}\mathbb{P}(\tau_{\textrm{second}} > s )\,\mathrm{d}s},
\end{equation}
where
\begin{equation}
\mathbb{P}(\tau_{\textrm{second}} > s ) = \mathbb{P}(\tau_1>s) + \mathbb{P}(\tau_2>s) - \mathbb{P}(\tau_1>s,\tau_2>s).
\end{equation}

\subsection{CDS Basket Results}\label{cdsresults}

\begin{figure}[htbp]
\setlength{\unitlength}{1cm}
\begin{minipage}[t]{8.0cm}
\caption{First-to-default CDS, varying T}\label{fig:first_cds_time}
\includegraphics[width=8cm]{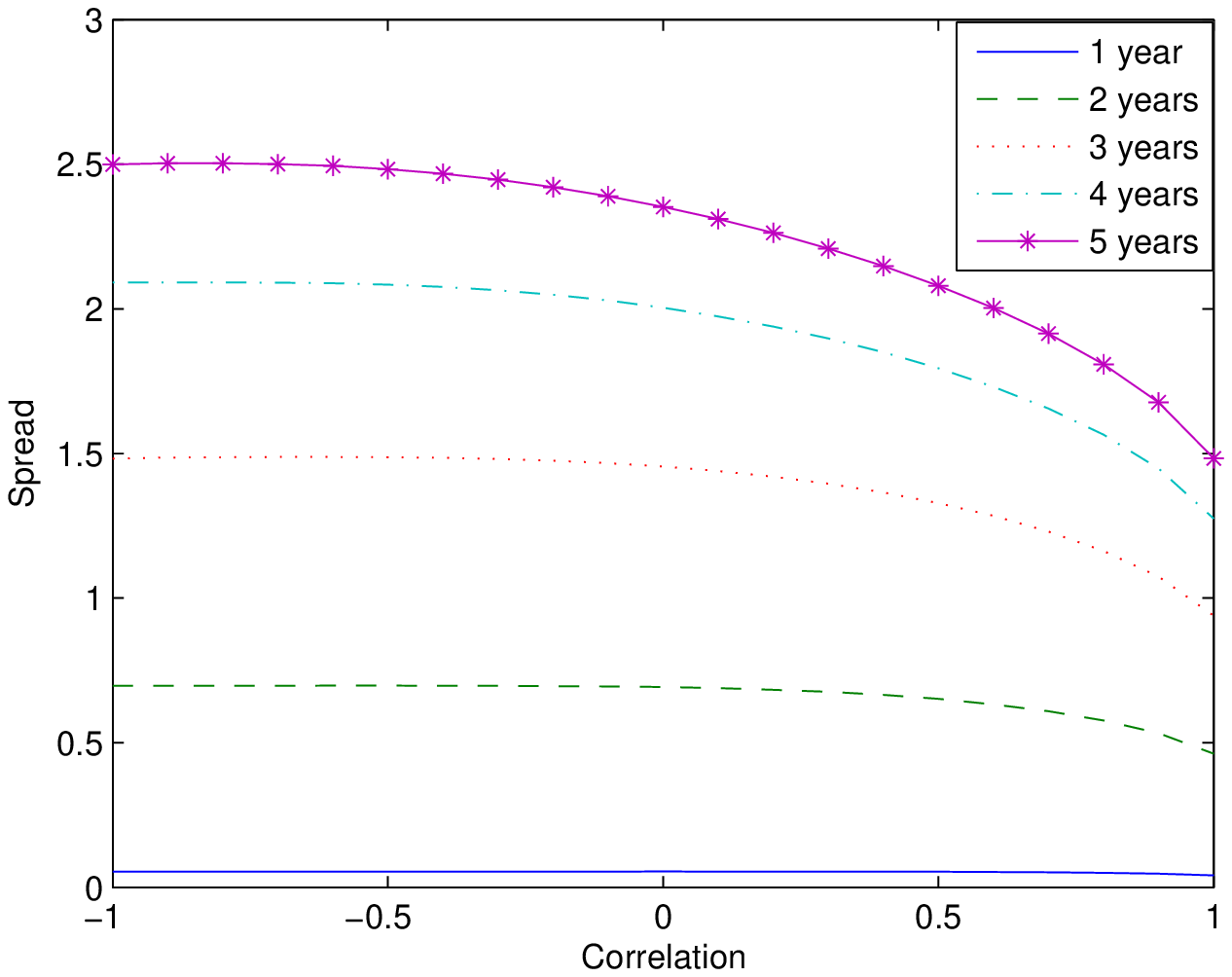} 
\end{minipage}
\hfill
\begin{minipage}[t]{8.0cm}
\caption{Second-to-default CDS, varying T}\label{fig:second_cds_time}
\includegraphics[width=8cm]{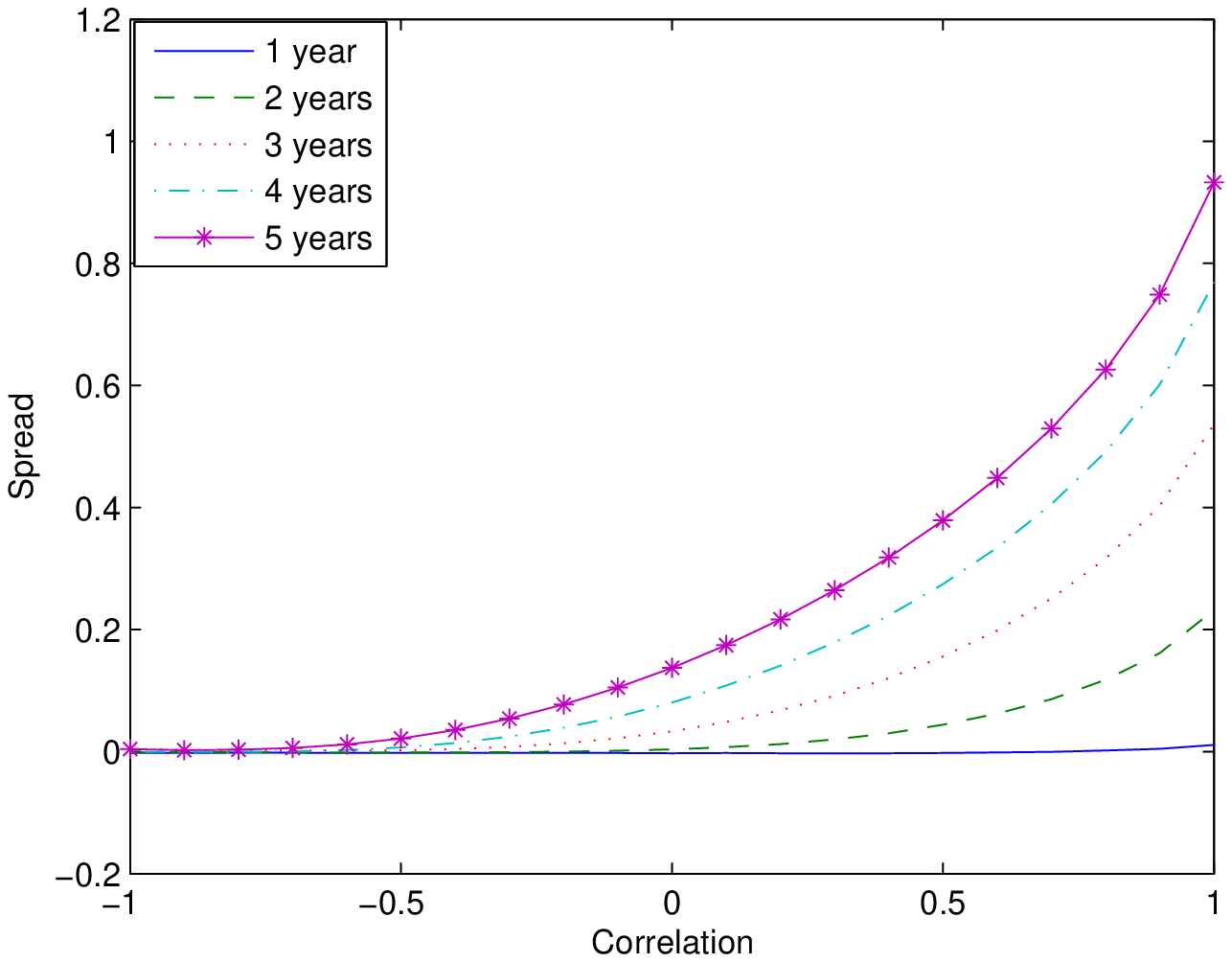}
\end{minipage}
\begin{center}
\small{$ \sigma_1=\sigma_2=0.2, K_1=100, r_{\!f}=0.05,q_1=q_2=0, $\\$ \gamma_1=\gamma_2=0.03$, initial credit quality = 2, $R=0.5$}
\end{center}
\end{figure}

\begin{figure}[htbp]
\setlength{\unitlength}{1cm}
\begin{minipage}[t]{8.0cm}
\caption{First-to-default CDS, varying $R$}\label{fig:first_cds_R}
\includegraphics[width=8cm]{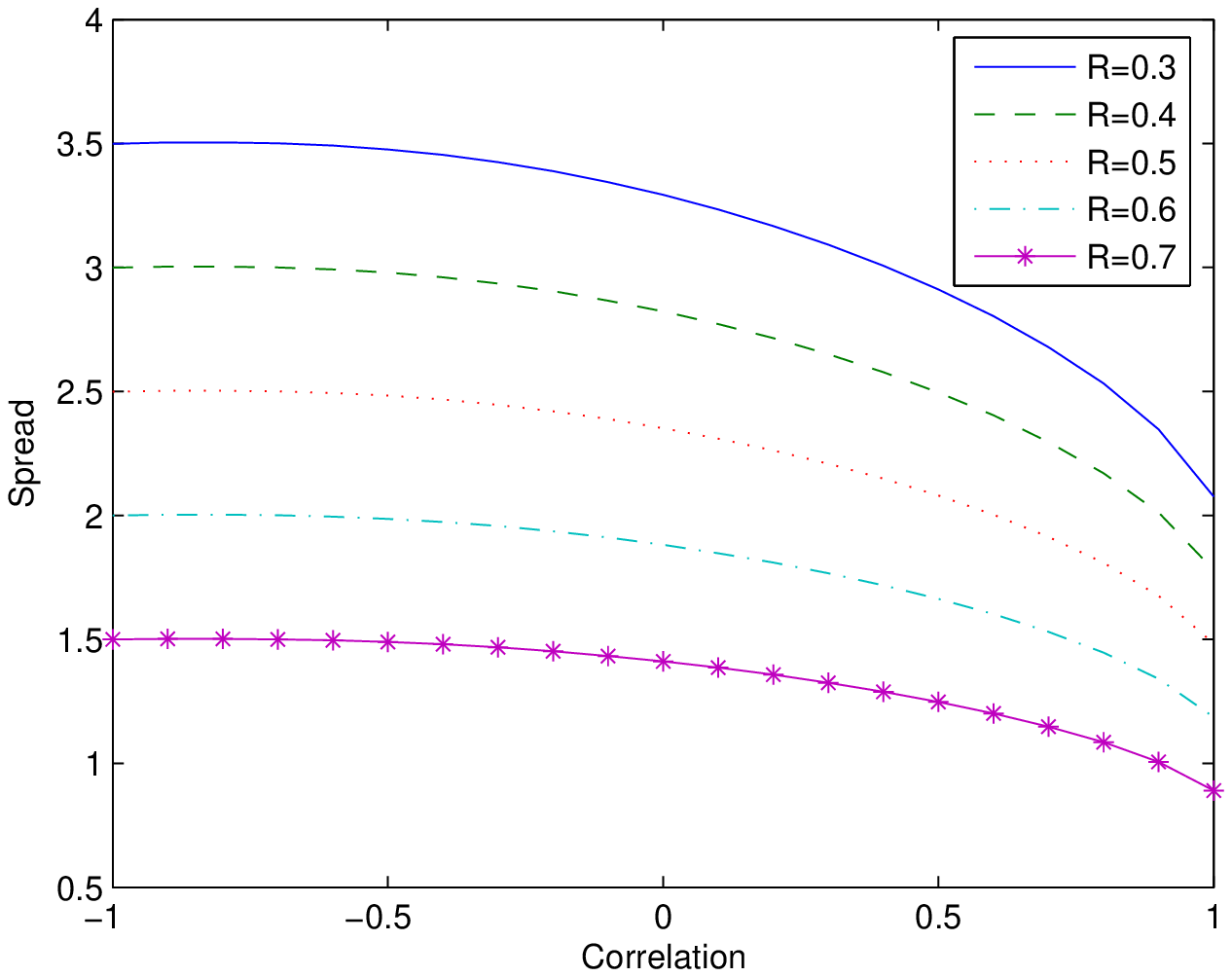} 
\end{minipage}
\hfill
\begin{minipage}[t]{8.0cm}
\caption{Second-to-default CDS, varying $R$}\label{fig:second_cds_R}
\includegraphics[width=8cm]{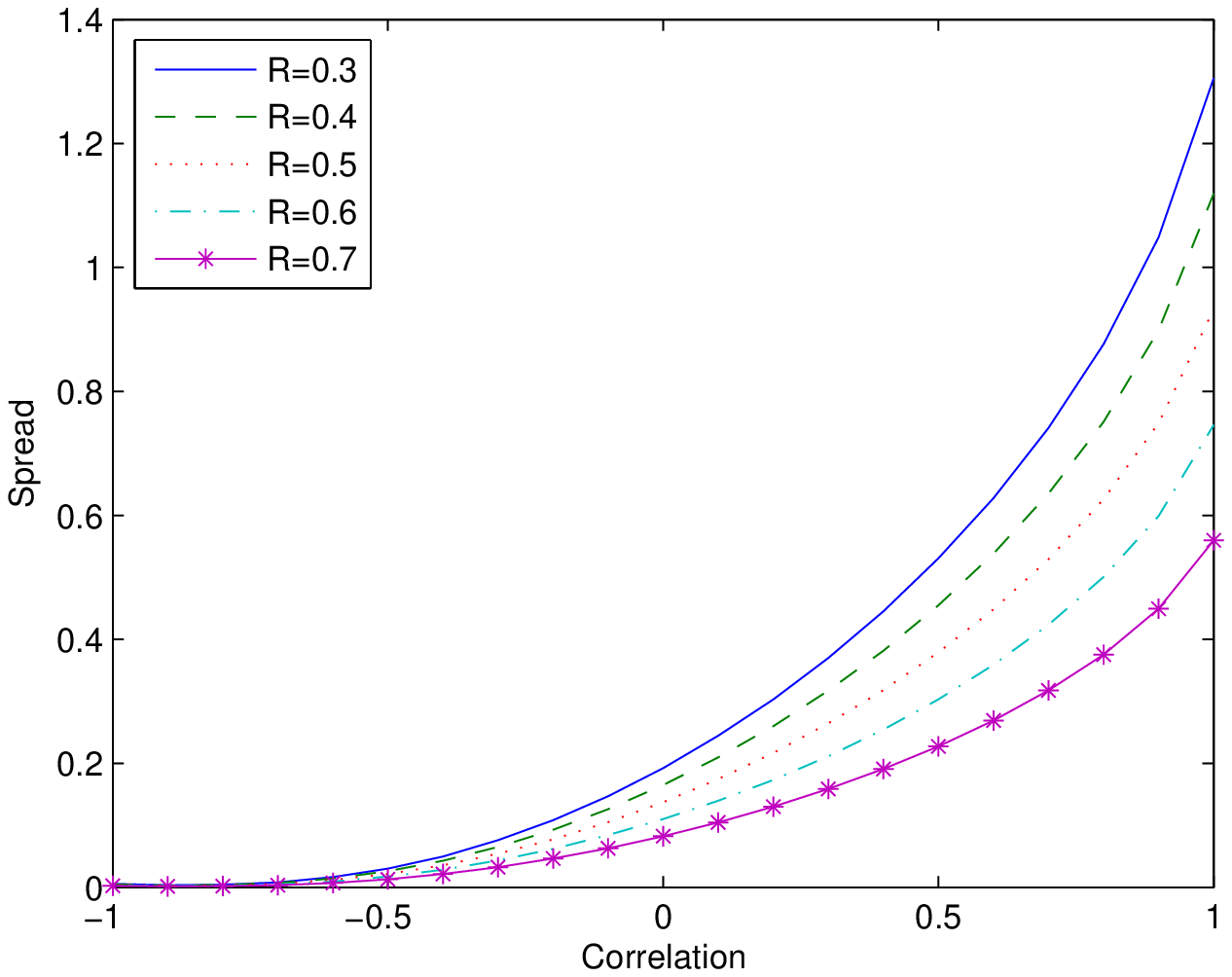}
\end{minipage}
\begin{center}
\small{$ \sigma_1=\sigma_2=0.2, K_1=100, r_{\!f}=0.05,q_1=q_2=0, $\\$ \gamma_1=\gamma_2=0.03$, initial credit quality = 2, T = 5}
\end{center}
\end{figure}

\begin{figure}[htbp]
\setlength{\unitlength}{1cm}
\begin{minipage}[t]{8.0cm}
\caption{First-to-default CDS, varying $\sigma_i$}\label{fig:first_cds_sigma}
\includegraphics[width=8cm]{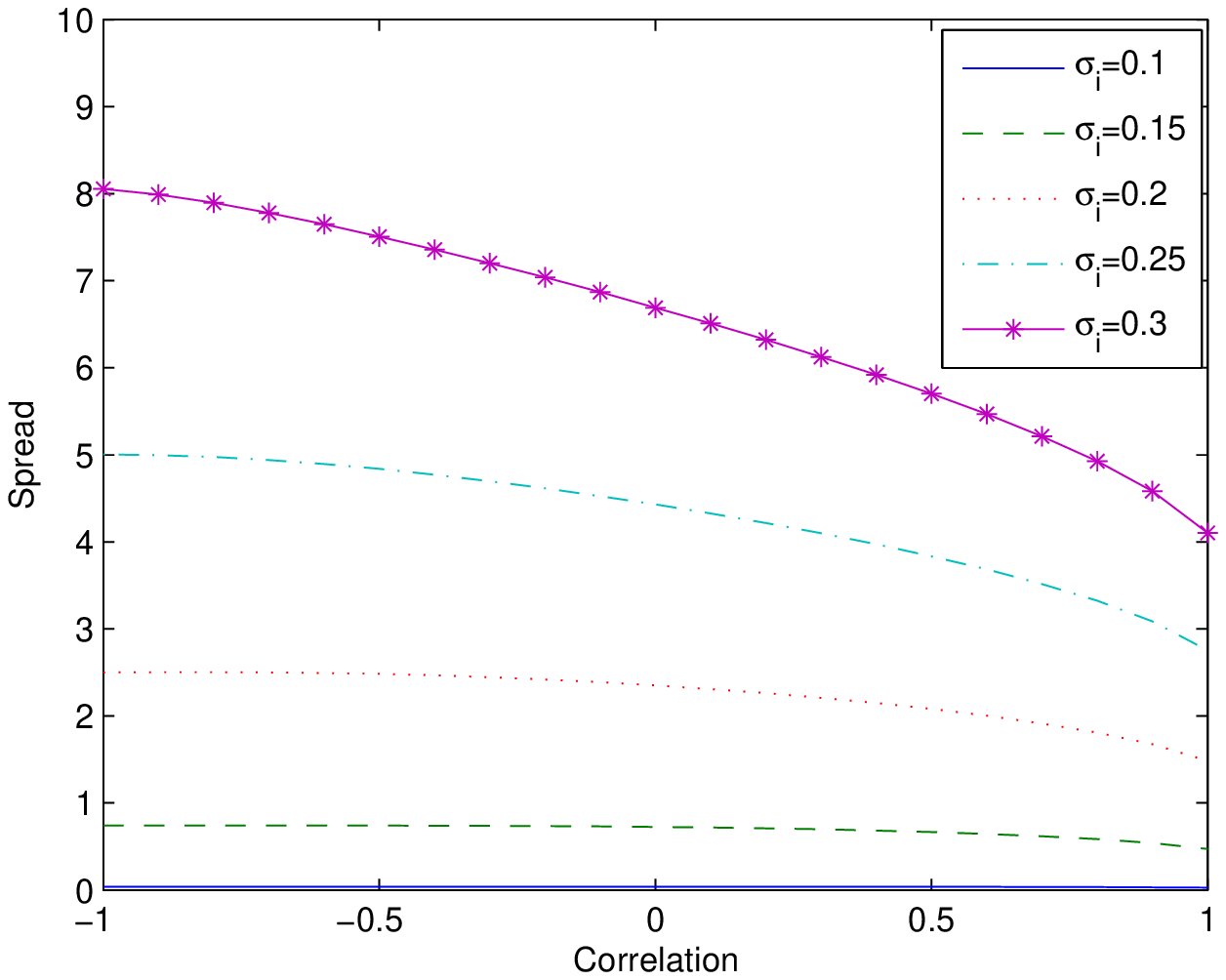} 
\end{minipage}
\hfill
\begin{minipage}[t]{8.0cm}
\caption{Second-to-default CDS, varying $\sigma_i$}\label{fig:second_cds_sigma}
\includegraphics[width=8cm]{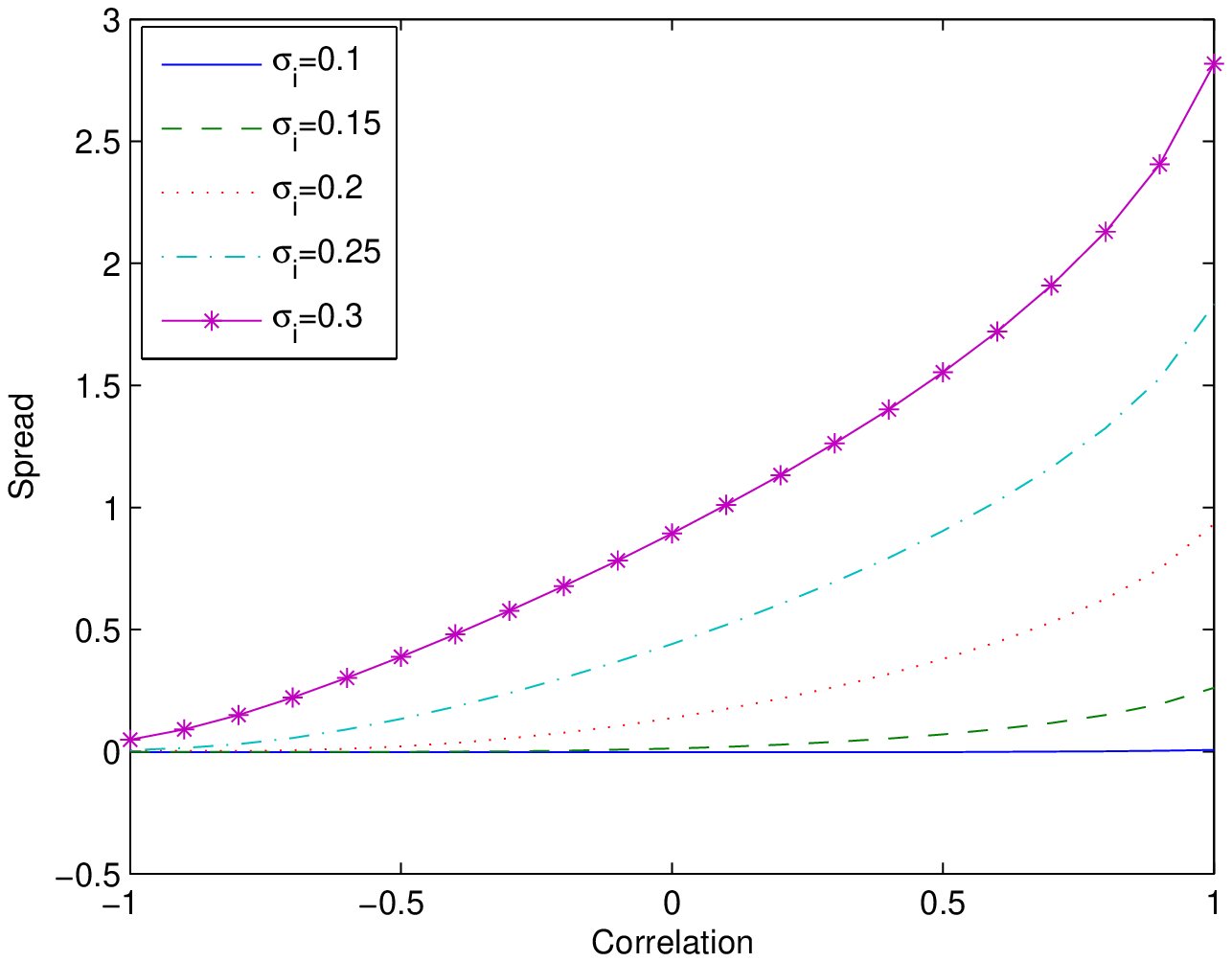}
\end{minipage}
\begin{center}
\small{$ K_1=100, r_{\!f}=0.05,q_1=q_2=0, R=0.5$\\$ \gamma_1=\gamma_2=0.03$, initial credit quality = 2, T = 5}
\end{center}
\end{figure}

In Figures \ref{fig:first_cds_time} - \ref{fig:second_cds_sigma}, we consider the impact of correlation on first and second-to-default CDS spreads for different parameter values. Numerical evaluation is done by numerical quadrature on a sparse grid as before. 

In all cases, with increasing correlation between the two reference entities, first-to-default CDS spreads decrease, whilst second-to-default CDS spreads increase. This is because the probability of at least one company defaulting in a given period is higher for negative correlations, whilst the probability of both defaulting is greater for positive correlations.

Figures \ref{fig:first_cds_time} and \ref{fig:second_cds_time} show spreads for first and second-to-default CDS baskets with maturities of up to 5 years. Initial credit quality is 2 (i.e, as before, firm value is initially twice the level of the barrier). Spreads are greater for longer-maturity swaps and, as we would expect, first-to-default spreads are everywhere greater than second-to-default spreads.

Figures \ref{fig:first_cds_R} and \ref{fig:second_cds_R} illustrate the extent to which CDS spreads depend on our recovery rate assumption. As would be expected, moving from a 30\% recovery rate to a 70\% recovery rate has a large impact. However, the overall form of spreads and their variation with changing correlation is the same. In general, taking R=50\% is representative of the levels seen in practice (see, for example, \cite{Bakshi-2006}) and is in line with that used in the CreditGrades$^{TM}$ approach to modelling credit as described in \cite{Finger-2002}. An easy extension would be to set the recovery rate of the reference entities equal to discounted par value.

Figures \ref{fig:first_cds_sigma} and \ref{fig:second_cds_sigma} show that firm volatility has a considerable impact on spreads. As the reference entities become more volatile, both credit default swaps become much more risky and spreads increase significantly. Of all the parameters, firm volatility has by far the largest impact on CDS spreads.

Through implementation of the analytical formula for the joint survival probability function, (\ref{eqn:survival}), we are therefore able to illustrate the sensitivity of first and second-to-default CDS spreads to input parameter assumptions straightforwardly. The importance of the degree of correlation between reference entities is clearly evident, with spreads significantly different when the basket is highly correlated than when it is well diversified.

Since by definition default contagion comes into play following the first default, its existence has no impact on first-to-default spreads. It is, however, important in the consideration of second-to-default spreads, and introducing the default contagion mechanism considered in Section \ref{bond} clearly has a dramatic impact on second-to-default spreads, in this case making them equal to first-to-default spreads. This situation is clearly extreme, as a CDS is highly unlikely to be constructed on such an undiversified basket (unless of course the existence of the link between firms was completely unknown prior to the first default), however it serves to highlight the importance of considering the dependence relationship between firms.

\subsection{CDS with Counterparty Risk}\label{counterparty}

Consider now a single-name CDS, face-value K, maturity T, on reference company one bought from a counterparty company two. The purchaser of the CDS makes spread payments for the life of the CDS -- until either the reference company or the counterparty defaults. If the reference entity defaults during the life of the CDS and before the counterparty, the purchaser receives a default payment. If, however, the counterparty defaults first, they receive nothing, irrespective of whether or not the reference company later defaults. Denoting the default time of company $i$ by $\tau_i$, if bond recovery on default is $R$ and the purchaser of protection makes continuous spread payments, $c$, for the life of the CDS, then using the same notation as before, the protection buyer 

\begin{itemize}
\item makes spread payments for $t < \min\{\tau_1, \tau_2, T\}$,
\item receives a default payment if $\tau_1 < \min\{\tau_2,T\}$,
\item receives nothing if $\tau_2 < \min\{\tau_1,T\}$.
\end{itemize}

In other words, discounted spread and default payments are 

\begin{eqnarray*}
\textrm{Spread } &=& cK\int_0^T e^{-r_{\!f}s}\mathbb{P}(\tau_1>s, \tau_2>s)\,\mathrm{d}s\\
&=& cK\int_0^T e^{-r_{\!f}s}\mathbb{P}(\underline{X}_1(s) \geq B_1,\underline{X}_2(s) \geq B_2)\,\mathrm{d}s\\
\textrm{Default } &=& (1-R)K \bigg\{\int_0^T e^{-r_{\!f}s}\mathbb{P}(s \leq \tau_1 \leq s+ ds, \tau_2 >s) \,\mathrm{d}s
\end{eqnarray*}

Considering the default payment for a first-to-default CDS, equation (\ref{cds1default}),

\begin{eqnarray}
\textrm{DDP} &=& (1-R)K \int_0^T e^{-r_{\!f}s}\mathbb{P}(s \leq \tau_{\textrm{first}} \leq s + ds )\,\mathrm{d}s\label{symmetry}\\
&=& (1-R)K \int_0^T e^{-r_{\!f}s}\bigg\{\mathbb{P}(s \leq \tau_1 \leq s+ ds, \tau_2>s)+\mathbb{P}(\tau_1>s,s \leq \tau_2\leq s+ ds)\bigg\}\,\mathrm{d}s,\nonumber
\end{eqnarray}
we see that the default payment for a CDS with counterparty, added to its image when the identity of the reference entity and the counterparty are swapped, gives the value of the first-to-default swap payment. A similar identity holds for the second-to-default swap payment. 

In the case of a homogeneous portfolio, when both reference entities have the same parameters, (\ref{symmetry}) can be used to calculate the value of a CDS spread with counterparty risk for all values of correlation, $\rho$. More generally, in the asymmetric case a more complicated evaluation must be done, details of which will be considered elsewhere. 

\section{Conclusion}\label{conclusion}

Structural models are increasingly the focus of industry attention in the multi-firm setting, but there has been limited work on the general first passage framework, and little academic coverage. This paper makes a first and novel, albeit by necessity simplistic, contribution to this field through the introduction of a contagion mechanism in a two-dimensional first passage model. Working with a \cite{Black-1976} type structural framework, we have built on the work by \cite{Zhou-2001b} to derive analytical formulae for both bond yields and CDS spreads. We have modified the default barrier to better reflect reality and have incorporated default contagion within the structural framework for the first time. The result is a credit model that is asymmetric with respect to default risk and which has a dependence structure based on both long-term asset correlation and default contagion.

Results illustrate that the sensitivity of yields to input parameters is as expected, and clearly demonstrate the importance of credit correlation. Our model enables us to generate corporate bond yields and CDS spreads across the full range of parameter values in a two-dimensional first passage framework in full generality. This has not been done before. Previous results using related analysis in \cite{Zhou-2001b} (default correlations) and \cite{Rebholz} (double lookbacks) have concentrated on cases in which the framework simplifies and have been limited to a few parameter values in these cases. For the first time, we are therefore able to fully consider spread sensitivity to model and parameter assumptions in the two-dimensional structural setting.

Our specification of default contagion is clearly not very realistic -- default by one company very rarely leads to direct default by another, although it is possible. More likely, the impact of a corporate bankruptcy causes a ripple of credit weakness through the market as related companies are impacted.\footnote{We address this numerically in a subsequent paper, \cite{Haworth2}, however analytical solutions are no longer possible.} Nonetheless, the importance of taking into account credit interactions is, once again, clearly highlighted.

Dependence modelling is most critical in the analysis and pricing of large basket credit derivatives, such as k$^{th}$-to-default credit default swap baskets and CDO tranches. These require the framework to be extended to considerably more than two dimensions. This is an area of current research interest and is not straightforward since analytical formulae are no longer possible and numerical solutions become highly problematic with increasing dimension. As intimated in Section \ref{bond}, it would be attractive to incorporate a network of asymmetric dependences within a portfolio, enabling the impact of a credit event at one company to cause a ripple of contagion through other, related, parties.

\appendix

\section{\label{app:1} Derivation of Maturity Payment}
For ease of notation, we denote the joint survival probability transition density
\begin{displaymath}
p(x_1, x_2, t) = \frac{\partial^2}{\partial x_1 \partial x_2}\mathbb{P}(X_1(t) \leq x_1, X_2(t) \leq x_2, \underline{X}_1(t) \geq B_1, \underline{X}_2(t) \geq B_2) 
\end{displaymath}

\begin{Proposition} \label{propAI}

For general $A$ and $B$,
\begin{eqnarray*}
\lefteqn{\int_{B_2}^{\infty}\int_A^B e^{\epsilon x_1}p(x_1,x_2,t)\,\mathrm{d}x_1 \,\mathrm{d}x_2 } \\
&& {}= \frac{2}{\beta t} e^{(a_1+\epsilon)B_1 + a_2B_2 + bt} \sum_{n=1}^{\infty} e^{-r_0^2/2t} \sin \left(\frac{n\pi\theta_0}{\beta}\right) \int_0^{\beta} \sin \left(\frac{n\pi\theta} {\beta}\right) g_n(\theta) \,\mathrm{d}\theta
\end{eqnarray*}
where
\begin{eqnarray*}
g_n(\theta) &=& \int_{d_A}^{d_B} r e^{-r^2/2t} e^{[A(\theta) + \epsilon\sigma_1\sin(\beta-\theta)]r} I_{(\frac{n\pi}{\beta})}\left(\frac{rr_0}{t}\right) \,\mathrm{d}r \\
r_0 &=& \frac{1}{\sqrt{1-\rho^2}}\left( \frac{B_1^2}{\sigma_1^2} - \frac{2\rho B_1 B_2}{\sigma_1\sigma_2} + \frac{B_2^2}{\sigma_2^2} \right)^{1/2} \\
\tan \theta_0 &=& \frac{\sigma_1 B_2 \sqrt{1-\rho^2}}{\sigma_2 B_1 - \rho \sigma_1 B_2}, \quad \theta_0 \in [0,\beta]\\
d_A &=& \frac{A-B_1}{\sigma_1\left[\sqrt{1-\rho^2}\cos \theta + \rho \sin \theta \right]} \\
d_B &=& \frac{B-B_1}{\sigma_1\left[\sqrt{1-\rho^2}\cos \theta + \rho \sin \theta\right]}\\
A(\theta) &=& a_1\sigma_1 \sin(\beta-\theta) + a_2\sigma_2\sin\theta,
\end{eqnarray*}
\end{Proposition}

\emph{Proof}

Using notation from Section \ref{model}, by \cite{Rebholz}
\begin{equation*}
p(x_1,x_2,t) = \frac{2e^{a_1x_1 + a_2x_2 + bt}}{\beta t\sigma_1\sigma_2\sqrt{1-\rho^2}} \sum_{n=1}^{\infty} e^{-(r^2+r_0^2)/2t} \sin \left(\frac{n\pi\theta_0}{\beta} \right)\sin \left( \frac{n\pi\theta}{\beta}\right) I_{(\frac{n\pi}{\beta})} \left( \frac{rr_0}{t} \right).
\end{equation*}
Changing variables,
\begin{eqnarray}\label{tranform}
x_1 &=& B_1 + \sqrt{(1-\rho^2)}\sigma_1 r \cos \theta + \rho \sigma_1 r \sin \theta \\
x_2 &=& B_2 + \sigma_2 r \sin \theta, \nonumber
\end{eqnarray}
the Jacobian for the transformation is $\sqrt{(1-\rho^2)} r \sigma_1 \sigma_2$, and 

\begin{eqnarray*}
\lefteqn{ \int_{B_2}^{\infty}\int_A^B e^{\epsilon x_1}p(x_1,x_2,t)\,\mathrm{d}x_1 \,\mathrm{d}x_2}  \\
&& {}= \frac{2}{\beta t} e^{(a_1+\epsilon)B_1 + a_2B_2 + bt} \sum_{n=1}^{\infty} e^{-r_0^2/2t} \sin \left(\frac{n\pi\theta_0}{\beta}\right) \int_{\theta} \sin \left(\frac{n\pi\theta} {\beta}\right) g_n(\theta) \,\mathrm{d}\theta
\end{eqnarray*}

where
\begin{displaymath}
g_n(\theta) = \int_r r e^{-r^2/2t} e^{[A(\theta) + \epsilon\sigma_1\sin(\beta-\theta)]r} I_{(\frac{n\pi}{\beta})}\left(\frac{rr_0}{t}\right) \,\mathrm{d}r, 
\end{displaymath}
since 
\begin{equation}
\cos \beta = -\rho \quad \& \quad \sin \beta = \sqrt{1-\rho^2}. \nonumber
\end{equation}

Writing $X=\frac{x_1-B_1}{\sigma_1}$ and $Y = \frac{x_2-B_2}{\sigma_2}$ then from (\ref{tranform}),
\begin{eqnarray}\label{XY}
X &=&  r\left[\sqrt{(1-\rho^2)}\cos \theta + \rho \sin \theta \right]\\
Y &=&  r \sin \theta \nonumber
\end{eqnarray}

Since
\begin{eqnarray*}
0 \leq \frac{A-B_1}{\sigma_1} &\leq& X \leq \frac{B-B_1}{\sigma_1} \\
0 &\leq& Y\leq \infty
\end{eqnarray*}

it follows that
\begin{displaymath}
\tan \theta = \frac{\sqrt{1-\rho^2}}{X/Y - \rho},\quad \Rightarrow \quad
\theta \in [0,\beta]
\end{displaymath}

and letting
\begin{eqnarray*}
d_A = \frac{A-B_1}{\sigma_1\left[\sqrt{1-\rho^2}\cos \theta + \rho \sin \theta \right]} \\
d_B = \frac{B-B_1}{\sigma_1\left[\sqrt{1-\rho^2}\cos \theta + \rho \sin \theta\right]},
\end{eqnarray*}
we have $d_A \leq r \leq d_B$ from (\ref{XY}). The result follows.

\begin{flushright}
$\Box$
\end{flushright}

\subsection*{Derivation of Maturity Payment}

From Section \ref{bond}, the discounted maturity payment, DMP, is:
\begin{eqnarray}\label{app1line1}
\textrm{DMP} &=& e^{-r_{\!f}T}  \int_{B_2}^{\infty} \int_d^{\infty} K_1 p(x_1, x_2, T) \,\mathrm{d}x_1 \,\mathrm{d}x_2\\\label{app1line2}
&& {}+  e^{-r_{\!f}T}  \int_{B_2}^{\infty} \int_{B_1}^d \omega_1 V_1(0) e^{x_1 + \gamma_1 T} p(x_1, x_2, T) \,\mathrm{d}x_1 \,\mathrm{d}x_2 
\end{eqnarray}

Using Proposition (\ref{propAI}) with $\epsilon=0$, $A=d$ and $B=\infty$ for line (\ref{app1line1}) and $\epsilon=1$, $A=B_1$ and $B=d$ for line (\ref{app1line2}), the payment on maturity becomes:

\begin{eqnarray*}
\textrm{DMP} &=& H_1(T) \sum_{n=1}^{\infty} \sin \left(\frac{n\pi\theta_0}{\beta}\right) \int_0^{\beta} \sin \left(\frac{n\pi\theta} {\beta}\right) g_n^+(\theta) \,\mathrm{d}\theta\\
&& {}+ H_2(T) \sum_{n=1}^{\infty} \sin \left(\frac{n\pi\theta_0}{\beta}\right) \int_0^{\beta} \sin \left(\frac{n\pi\theta} {\beta}\right) g_n^{*}(\theta) \,\mathrm{d}\theta
\end{eqnarray*}
where,
\begin{eqnarray*}
H_1(T) &=& \frac{2K_1e^{-r_{\!f}T}}{\beta T} e^{a_1B_1 + a_2B_2 + bT} e^{-r_0^2/2T}\\
H_2(T) &=& \frac{2\omega_1V_1(0)e^{(\gamma_1-r_{\!f})T}}{\beta T} e^{(a_1+1)B_1 + a_2B_2 + bT} e^{-r_0^2/2T}\\
g_n^+(\theta) &=& \int_{d^*(\theta)}^{\infty} r e^{-r^2/2T} e^{A(\theta)r} I_{(\frac{n\pi}{\beta})}\left(\frac{rr_0}{T}\right) \,\mathrm{d}r \\
g_n^{*}(\theta) &=& \int_{0}^{d^*(\theta)} r e^{-r^2/2T} e^{[A(\theta) + \epsilon\sigma_1\sin(\beta-\theta)]r} I_{(\frac{n\pi}{\beta})}\left(\frac{rr_0}{T}\right) \,\mathrm{d}r \\
d^*(\theta) &=& \frac{d-B_1}{\sigma_1\left[\sqrt{1-\rho^2}\cos \theta + \rho \sin \theta \right]}\\
&=& \frac{\ln \omega_1}{\sigma_1\sin(\theta-\beta)} \geq 0.
\end{eqnarray*}


\bibliography{2DArticle_new.tex}
\bibliographystyle{plainnat}
\end{document}